\documentclass[11pt, a4paper]{article}

\usepackage{latexsym, graphicx}
\usepackage{amsmath, amssymb, amsthm}
\usepackage{hyperref} 

\newtheorem{theorem}{Theorem}

\newtheorem{observation}[theorem]{Observation}

\theoremstyle{definition}

\newcommand\dist{{\rm dist}}

\newcommand{\ZZ}{\mathbb{Z}^2}

\begin{document}

\title{The packing chromatic number of the square lattice is at least 12
\thanks{This work was supported by the Ministry of Education of the Czech Republic as projects 1M0021620808, 1M0545 and GACR~201/09/0197.
}
}

\author{
Jan Ekstein $^a$ 
\and
Ji\v{r}\'{\i} Fiala $^b$ 
\and
P\v{r}emysl Holub $^a$ 
\and 
Bernard Lidick\'y $^b$}

\date{\today}

\maketitle

\begin{center}{\small
$^a$
Department of Mathematics and Inst. for Theoretical Computer Science (ITI), 
University of West Bohemia, Univerzitn\'{\i} 22, 306 14 Pilsen, Czech Republic\\
E-mail: {\tt \{ekstein,holubpre\}@kma.zcu.cz}
\medskip

$^b$
Department of Applied Mathematics and Inst. for Theoretical Computer Science (ITI),\\
Charles University, Malostransk\'e n\'am. 25, 118 00 Prague, Czech Republic\\
E-mail: {\tt \{fiala,bernard\}@kam.mff.cuni.cz}
\medskip
}
\end{center}

\begin{abstract}
The packing chromatic number $\chi_\rho(G)$ of a graph $G$ is the 
smallest integer $k$ such that the vertex set $V(G)$ can be partitioned into disjoint
classes $X_1, \ldots, X_k$, where vertices in $X_i$ have 
pairwise distance greater than $i$.

For the 2-dimensional square lattice $\ZZ$
it is proved  that $\chi_\rho(\ZZ) \geq 12$,
which  improves the previously known lower bound 10.
\end{abstract}


\noindent
{\bf ACM 1998 classification }: G.2.2 Graph theory


\section{Introduction}

The concept of packing coloring comes from the area of frequency planning in wireless
networks. This model emphasizes the fact that some frequencies have
higher throughput and hence they are used more sparely to avoid
an interference. 

In our model, the first frequency cannot be assigned to neighbouring
nodes. The second frequency cannot be assigned to nodes in distance at most two
and so on. 
In graph theory language we ask for a partition of the vertex set of a
graph $G$ into disjoint color classes $X_1,\ldots,X_k$ according to the 
following constraints. Each color class $X_i$ should be an \emph{$i$-packing}, 
that is, a set of vertices
with the property that any distinct pair $u,v\in X_i$ satisfies $\dist(u,v)>i$. 
Here $\dist(u,v)$ denotes the shortest path distance between $u$ and $v$. 
Such partition is called a \emph{packing $k$-coloring},
even though it is allowed that some sets $X_i$ may be empty.
The smallest integer $k$ for which there exists a packing $k$-coloring of
$G$ is called the \emph{packing chromatic number} of $G$ and it is
denoted by $\chi_\rho(G)$. 

This concept, under the notion \emph{broadcast chromatic number}, was introduced by
Goddard et al.~\cite{gohe-08}. The notion packing chromatic was
proposed by Bre\v{s}ar et al.~\cite{brklra-07}.

Topic of this work is the packing chromatic number of the infinite
square lattice $\ZZ$. The question of determining $\chi_\rho(\ZZ)$
was posed in \cite{gohe-08}. Also a lower bound 9 and an upper bound 23
were given there. The upper bound was improved by Schwenk~\cite{schwenk-02} to 22 and later
by Holub and Soukal~\cite{holub-09} to 17.
The lower bound was improved to 10 by Fiala et al. ~\cite{fili-09}.

We further improve the lower bound from 10 to 12. 

\begin{theorem}\label{thm:main}
The packing chromatic number for the square lattice is at least 12.
\end{theorem}

The proof relies on computer. In the next section we describe the main idea
of the algorithm, which proves the theorem. All necessary code for
running the computation is available at
\url{http://kam.mff.cuni.cz/~bernard/packing}.

\section{The Result}

The algorithm for proving Theorem~\ref{thm:main} is a brute force search through all possible 
configurations on lattice $15\times9$. It is too time consuming to simply check everything. 
Hence we use the following observation to speed up the computation
by avoiding several configurations. 

\begin{observation}
If there exists a coloring of the square lattice with 11 colors then it is possible
to color lattice $15\times9$ where color 9 is at position $[5,5]$.
\end{observation}

Any other color at any other position could be fixed instead of 9 at $[5,5]$.
Color 9 at $[5,5]$ just sufficiently reduces the number of configurations to check.
We do not claim that it is the optimal choice.

If there exists a coloring we simply find any vertex of color 9 and take a piece
of the lattice in its neighborhood. 

So in the search through the configurations we assume that at position $[5,5]$ is
precolored by 9.
The coloring procedure gets a matrix and tries to color the vertices
row by row. A pseudocode is given here in this note.

\newpage

\begin{verbatim}
function boolean try_color(lattice, [x,y])
begin 
  for color := 1 to 11 do
     if can use color on lattice at [x,y] then
        lattice[x,y] := color
        if [x,y] is the last point return true
        else if try_color(lattice, next([x,y]) then return true
     endif
  endfor
  return false
end
\end{verbatim}

We have two implementations of this procedure. One is in the language C++ and
the other is in Pascal. The first one
is available online at \url{http://kam.mff.cuni.cz/~bernard/packing}.
We include the full source code as well as descriptions of inputs and
outputs. We were checking the outputs of both programs during the computation 
and we verified that they match. The total number of checked configurations was
43112312093324. The computation took about 120 days of computing time on a single core
workstation in year 2009.

The procedure fails to color the matrix $15 \times 9$ with 9 at position $[5,5]$.
Hence we conclude that the packing chromatic number for the square lattice
is at least 12.

\end{document}